\documentclass[sigconf]{acmart}

\usepackage{balance}
\usepackage{subcaption}
\usepackage{listings}
\usepackage{cleveref}
\usepackage{todonotes}

\usepackage{enumitem}

\usepackage{tikz}
\usetikzlibrary{shadows,calc}

\def\shadowshift{2pt,-2pt}
\def\shadowradius{6pt}

\colorlet{innercolor}{gray!60}
\colorlet{outercolor}{gray!05}

\newcommand\drawshadow[1]{
    \begin{pgfonlayer}{shadow}
        \shade[outercolor,inner color=innercolor,outer color=outercolor] ($(#1.south west)+(\shadowshift)+(\shadowradius/2,\shadowradius/2)$) circle (\shadowradius);
        \shade[outercolor,inner color=innercolor,outer color=outercolor] ($(#1.north west)+(\shadowshift)+(\shadowradius/2,-\shadowradius/2)$) circle (\shadowradius);
        \shade[outercolor,inner color=innercolor,outer color=outercolor] ($(#1.south east)+(\shadowshift)+(-\shadowradius/2,\shadowradius/2)$) circle (\shadowradius);
        \shade[outercolor,inner color=innercolor,outer color=outercolor] ($(#1.north east)+(\shadowshift)+(-\shadowradius/2,-\shadowradius/2)$) circle (\shadowradius);
        \shade[top color=innercolor,bottom color=outercolor] ($(#1.south west)+(\shadowshift)+(\shadowradius/2,-\shadowradius/2)$) rectangle ($(#1.south east)+(\shadowshift)+(-\shadowradius/2,\shadowradius/2)$);
        \shade[left color=innercolor,right color=outercolor] ($(#1.south east)+(\shadowshift)+(-\shadowradius/2,\shadowradius/2)$) rectangle ($(#1.north east)+(\shadowshift)+(\shadowradius/2,-\shadowradius/2)$);
        \shade[bottom color=innercolor,top color=outercolor] ($(#1.north west)+(\shadowshift)+(\shadowradius/2,-\shadowradius/2)$) rectangle ($(#1.north east)+(\shadowshift)+(-\shadowradius/2,\shadowradius/2)$);
        \shade[outercolor,right color=innercolor,left color=outercolor] ($(#1.south west)+(\shadowshift)+(-\shadowradius/2,\shadowradius/2)$) rectangle ($(#1.north west)+(\shadowshift)+(\shadowradius/2,-\shadowradius/2)$);
        \filldraw ($(#1.south west)+(\shadowshift)+(\shadowradius/2,\shadowradius/2)$) rectangle ($(#1.north east)+(\shadowshift)-(\shadowradius/2,\shadowradius/2)$);
    \end{pgfonlayer}
}

\pgfdeclarelayer{shadow} 
\pgfsetlayers{shadow,main}

\newcommand\shadowimage[2][]{%
\begin{tikzpicture}
\node[anchor=south west,inner sep=0] (image) at (0,0) {\includegraphics[#1]{#2}};
\drawshadow{image}
\end{tikzpicture}}

\copyrightyear{2023}
\acmYear{2023}
\setcopyright{rightsretained}
\acmConference[UIST '23]{The 36th Annual ACM Symposium on User Interface Software and Technology}{October 29-November 1, 2023}{San Francisco, CA, USA}
\acmBooktitle{The 36th Annual ACM Symposium on User Interface Software and Technology (UIST '23), October 29-November 1, 2023, San Francisco, CA, USA}
\acmDOI{10.1145/3586183.3606819}
\acmISBN{979-8-4007-0132-0/23/10}

\begin{document}

\title{Odyssey: An Interactive Workbench for Expert-Driven Floating-Point Expression Rewriting}
\author{Edward Misback}
\affiliation{%
  \institution{University of Washington}
  \city{Seattle}
  \state{Washington}
  \country{United States}
}
\email{misback@cs.washington.edu}
\author{Caleb C. Chan}
\affiliation{%
  \institution{University of Washington}
  \city{Seattle}
  \state{Washington}
  \country{United States}
}
\email{calebcha@cs.washington.edu}
\author{Brett Saiki}
\affiliation{%
  \institution{University of Washington}
  \city{Seattle}
  \state{Washington}
  \country{United States}
}
\email{bsaiki@cs.washington.edu}
\author{Eunice Jun}
\affiliation{%
  \institution{University of Washington}
  \city{Seattle}
  \state{Washington}
  \country{United States}
}
\affiliation{%
  \institution{University of California, Los Angeles}
  \city{Los Angeles}
  \state{California}
  \country{United States}
}
\email{emjun@cs.washington.edu}
\author{Zachary Tatlock}
\affiliation{%
  \institution{University of Washington}
  \city{Seattle}
  \state{Washington}
  \country{United States}
}
\email{ztatlock@cs.washington.edu}
\author{Pavel Panchekha}
\affiliation{%
  \institution{University of Utah}
  \city{Salt Lake City}
  \state{Utah}
  \country{United States}
}
\email{pavpan@cs.utah.edu}

\begin{abstract}
  In recent years, researchers have proposed a number of automated tools to
  identify and improve floating-point rounding error in mathematical expressions.
However, users struggle to effectively apply these tools.
In this paper, we work with 
  novices, experts, and tool developers to 
  investigate user needs during the expression rewriting process.
We find that users follow an iterative design process.
They want to compare expressions on multiple input ranges, 
  integrate and guide various rewriting tools,
  and understand where errors come from.
We organize this investigation's results into a three-stage workflow
  and implement that workflow
  in a new, extensible workbench dubbed Odyssey.
Odyssey enables users to:
  (1) \textit{diagnose} problems in an expression, 
  (2) \textit{generate solutions} automatically or by hand, and 
  (3) \textit{tune} their results. 
Odyssey tracks a working set of expressions
  and turns a state-of-the-art automated tool ``inside out,''
  giving the user access to internal heuristics, algorithms, 
  and functionality.
In a user study, Odyssey enabled five expert numerical analysts
   to solve challenging rewriting problems
  where state-of-the-art automated tools fail.
In particular, the experts unanimously praised Odyssey’s novel support for
  interactive range modification and local error visualization.
\end{abstract}



\maketitle

\newcommand{\eunice}[1]{\textcolor{olive}{#1}}

\newcommand{\shortquote}[1]{``\emph{#1}''}
\newcommand{\longquote}[1]{\vspace{-1pt}\begin{quote}``\emph{#1}''\end{quote}}

\section{Introduction}

\begin{figure}
  \centering
  \begin{subfigure}[b]{0.4\textwidth} 
    \begin{lstlisting}[language=Python,commentstyle=\color{blue}] 
      # FP arithmetic seems ok
      >>> x = 1e15
      >>> x + 1 - x
      1.0

      # ... until it doesn't!
      >>> x = 1e16
      >>> x + 1 - x
      <@\textcolor{red}{\large 0.0}@>
    \end{lstlisting}
    \caption{
      Example of ``catastrophic cancellation'' in Python. \\[8pt]
    }
    \label{fig:cancellation}
  \end{subfigure}
  \hfill
  \begin{subfigure}[b]{0.45\textwidth}

    \includegraphics[width=\textwidth]{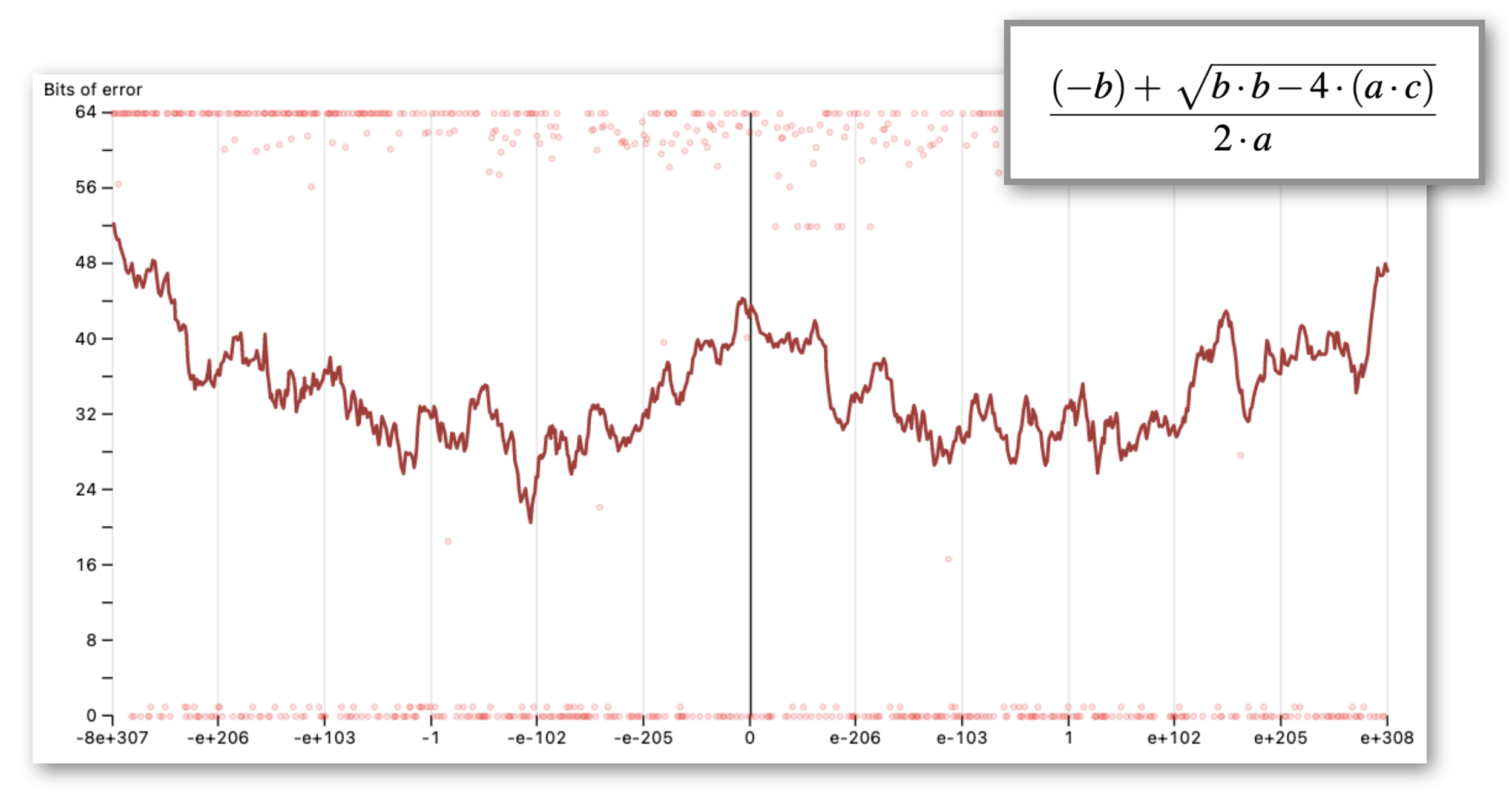}
    \caption{
      Average ``bits'' 
        ($\log_2(\mathit{ulps})$)  
        of floating-point error
        with respect to $b$
        when evaluating the quadratic equation over
        randomly-sampled inputs.
      For many applications the error is unacceptable,
        but few programmers are equipped
        to address such numerical issues.
    }
    \label{fig:quadp-error}
  \end{subfigure}
  \caption{
    Floating-point error is pernicious;
      even familiar, simple expressions
      can yield meaningless results.
  }
\end{figure}

Floating-point arithmetic is widely used
  in scientific, engineering, and graphical applications
  to approximate arithmetic on real numbers;
  typically, it is the
  only practical option available.%
\footnote{
  While alternatives exist,
    e.g., arbitrary-precision arithmetic,
    exact rational arithmetic,
    and constructive real arithmetic,
    they are orders of magnitude slower
    than hardware floating-point,
    and are thus inappropriate for many applications.
}
However, floating-point arithmetic must be used with care,
  as rounding errors can cause floating-point arithmetic
  and real-number arithmetic to give dramatically different results.
For example,
  naïve implementations of
  well-known mathematical equations like the quadratic formula
  can exhibit unacceptably-high rounding error (\Cref{fig:quadp-error}).
Rounding error can also ruin results
  for even extremely simple expressions.
\Cref{fig:cancellation} shows that,
  for large floating-point values of \texttt{x},
  the expression \texttt{x + 1 - x} can evaluate to 0
  instead of the mathematically correct 1!
Floating-point rounding error has caused
  unreproducible scientific research,
  distorted stock market indices,
  and wartime casualties~\cite{num-replication,distort-stock,patriot,euro-rounding,wall-street-distort-stock,num-issues-in-stat,round-elections}.

As a specific example, 
  a major bug in the implementation of \textit{asinh}/\textit{acosh} 
  in the Rust standard math library went unnoticed for seven years.
An automated test suite caught the bug in 2022 \cite{herbie-rust}.

In order to diagnose and repair this kind of error,
  \textit{numerical analysis experts} have developed
  techniques and tools for analyzing and rewriting 
  floating-point expressions over the last decade.
These tools support and facilitate automated
  test generation ~\cite{s3fp},
  error analysis ~\cite{gappa,vmcai11-fluctuat,daisy,satire,fm15-fptaylor},
  and repair ~\cite{salsa,herbie}.
For example,
  the open-source, state-of-the-art Herbie tool~\cite{herbie} takes as input a floating-point expression
  and uses algebraic and analytic identities to rewrite the expression
  via a complex search process.
Despite wide adoption of tools like Herbie in
  industrial and national labs,
  users still find results are too complicated and that
  tools overlook seemingly obvious rewritings.

\lstset{escapeinside={<@}{@>}}

Fully understanding and fixing the bug in Rust required
  rewriting the naive definition $log(x + sqrt(x^2 + 1))$ as
  $log1p(x + x/(hypot(1, x) + x))$.
To arrive at the solution,
  numerical analysts needed to 
  repurpose internal operations of existing tools
  and apply their own expert knowledge.
\textbf{This example illustrates how experts must work with a constellation 
  of complicated analysis tools,
  none of which answer their questions 
  about an expression directly.}
Our goal is to enable numerical analysis experts and 
  developers of mathematical libraries 
  to find and fix similar bugs and prevent their occurence in the future.

Towards this goal, we
  observed novices and experts in an in-lab design study and
  found that users struggle with
  specifying their objectives and interpreting Herbie's results,
  facing issues of tool/user objective mismatch,
  lack of trust in the automated tool,
  and a need for independent exploration.
We also identified a three-stage
  floating-point rewriting workflow:
  (1) \textit{diagnosing problems},
    in which users identify the problematic operations within expressions;
  (2) \textit{generating solutions},
    in which users gather potential expression rewritings
    from automated tools, references, or their own creativity;
  and (3) \textit{tuning},
    where users test, tweak, and compare different rewritings
    to optimize the resulting expression
    for their own accuracy, performance, and maintainability needs.
This workflow is not well-addressed by existing tools. 
For example, 
  end-to-end tools like Herbie 
  can take minutes to return a batch of analysis results, 
  and there is no tool support for comparing
  and improving rewritings
  drawn from multiple sources.

To support this workflow, we designed and implemented Odyssey, 
  an interactive workbench that
  allows users to identify problem areas in floating-point 
  expressions using error visualizations, collect and manage 
  expression rewrites using an interactive table, and
  combine rewrites to minimize rounding error.
Odyssey leverages Herbie as an analysis and rewriting engine
  but retains context about the user's objectives, allowing it
  to return common analyses in less than a second.

To evaluate the effectiveness of Odyssey,
  we conducted a study with five experts
  in numerical computing and floating-point arithmetic.
On average,
  the experts successfully completed
  five out of seven challenging tasks
  drawn from real-world numerical problems
  in roughly 40 minutes after a 12-minute tutorial.
The interactive nature of Odyssey
  enabled experts to concentrate on high-level problem-solving
  and facilitated the swift evaluation and comparison of expression rewritings. 

  Odyssey contributes to a growing body of work
  on expert tools. Unlike end-users, experts have highly specialized workflows and significant low-level implementation knowledge they need to express and incorporate in tools. 
  Examples of expert tools include Roly-poly, a tool for guided optimization of Halide image processing code~\cite{ikarashi2021guidedOptimization}; 
  PerformanceHat, a tool for analyzing application runtime performance~\cite{cito2018performanceHat}; 
  and Tsugite, a tool for interactive design and fabrication of wood joints
  designed for expert machinists with limited experience working in a particular domain~\cite{larsson2020tsugite}.
By combining the power of automated systems
  with a dynamic, human-driven workflow,
  Odyssey is an example of how to enable more users
  to work efficiently along-side automated tools
  in complex domains beyond floating-point.

This paper makes four contributions:
\begin{enumerate}
  \item An investigation of the needs of novices and experts,
    summarized in a three-stage workflow
    for floating-point expression rewriting:
    diagnosis, solution generation, and tuning.
    This workflow combines both automated tools and human rewritings.
  \item An iteratively developed workbench, Odyssey, 
    that supports this workflow.
  \item A study of Odyssey's effectiveness based on feedback from expert users
    who completed a set of challenging tasks
    drawn from real-world numerical problems.
  \item A discussion of the implications of our work
    for the design of interactive expert tools
    that combine human and automated design space search.
\end{enumerate}

\section{Background and Related Work}
Odyssey draws on techniques from the developer tool literature on program visualization and program history
  to addresses key challenges developers face
  in the domain of floating-point arithmetic.

\subsection{Program Visualization for Debugging}

Floating-point error analysis and repair
  involves a mix of debugging and performance optimization work.
Odyssey is thus inspired by work aimed at
  program visualization for debugging.
Systems such as Whyline~\cite{ko2004designing}, Timelapse~\cite{burg2013interactive}, and FireCrystal~\cite{oney2009firecrystal},
  which connect code with runtime behavior by visualizing execution traces,
  inspire several of Odyssey's interactions,
  including the interactive ``local error'' heatmap 
  visualizing per-operation floating-point error for a particular input.
Moreover,
  a series of papers on integrating visualizations with code,
  such as Theseus~\cite{lieber2014theseus}, 
  which provides always-on visualizations of runtime state;
  Projection Boxes~\cite{lerner2020projectionBoxes}, 
  which gives programmers more control
  over which runtime values are visualized;
  and Hoffswell et al.~\cite{hoffswell2018augmentingCodeWithInSituVis}, 
  which provides recommendations
  for embedding visualizations in code,
  are reflected in our design of Odyssey's error graph,
  which allows programmers to visualize floating-point error
  and control which input values and rewritings
  are visualized.
Odyssey sees similar benefits from these designs as prior work:
  opening up space for programmer exploration and observation,
  and thereby giving programmers a fuller understanding of the problem space
  and a richer set of interactions for comparison and repair.

That said, floating-point rounding error
  is a continuous, numeric quality of a program,
  and the ``tuning'' stage of numerical work therefore has
  a lot of analogs to performance optimization.
Beck et al.~\cite{beck2013perfBottlenecksVis} and PerformanceHat~\cite{cito2018performanceHat}, for example,
  visualize the proportion of runtime
  spent at each each line of code in the program. 
  These approaches inspire our ``heatmap'' design for local error information,
  coloring each floating-point operation in the program
  based on the amount of floating-point rounding error
  it contributes to the result.
The Roly-poly~\cite{ikarashi2021guidedOptimization} project is also quite similar to Odyssey,
  aiding developers in exploring and selecting performance optimizations
  for image processing code.
Odyssey explores a similar system-aided optimization workflow,
  but for accuracy instead of performance optimization.

\subsection{Maintaining and Reviewing Code Versions}

To understand, experiment with, and collaborate on code,
  developers author and compare
  multiple program alternatives and histories~\cite{codoban2015softwareHistory}.
Tools such as Azurite~\cite{yoon2015azurite}, Verdant~\cite{kery2018verdant}, and Variolite~\cite{kery2017variolite}
  provide explicit support for multiple program versions.
For example, Verdant helps data scientists
  compare, replay, and simplify histories
  for code in computational notebooks~\cite{kery2018verdant}. 
  Also, Head et al.~\cite{head2019managingMesses} introduce ``code gathering'' techniques
  that find the minimal code slices in a program
  that produce a selected set of results. 
Comparing and combining multiple alternative rewritings
  is a also key part of floating-point error repair.

Odyssey maintains a history of rewritings
  both to provide a history of how a rewriting was developed
  and also allow developers to visualize, compare, and combine
  multiple alternatives, providing explicit internal support
  to what would otherwise be internal mental operations,
  thereby reducing cognitive load and allowing developers
  to focus on the higher-level problem-solving aspects.

\subsection{Floating-Point Arithmetic and Numerical Analysis}

Floating-point arithmetic,
  defined by the IEEE~754 standard~\cite{ieee08-standard},
  and variations of this standard
  form the standard number representation
  in most programming languages~\cite{toplas08-pitfalls-verifying}.
However, floating-point arithmetic is subject to rounding error,
  and even elementary computations often permit significant
  error~\cite{acm91-every-scientist}.
Numerical analysis provides a set of mathematical tools
  to analyze, bound, and reduce this error~\cite{book87-nmse}.
However, many programmers are unfamiliar with
  numerical analysis techniques, and even fewer have
  a thorough understanding of how to apply these tools.

Researchers have thus developed a vast menagerie of tools
  automating specific numerical analysis techniques,
  including Rosa~\cite{popl14-rosa} for affine arithmetic,
  FPTaylor~\cite{fm15-fptaylor} for error Taylor series, and
  Ariadne~\cite{popl13-ariadne} for root finding.
Other tools repurpose static analysis techniques
  to find floating-point rounding errors;
  such tools include Fluctuat~\cite{vmcai11-fluctuat},
  which uses abstract interpretation;
  FPDebug~\cite{pldi12-fpdebug},
  which uses a dynamic execution with shadow variables;
  and CGRS~\cite{ppopp14-cgrs}, which uses evolutionary search. 
These tools can find inputs with high rounding error
  or, in some cases, certify the absence of such errors.
Programmers can then use the error found
  to attempt to understand the source of the rounding error,
  and ultimately fix it.
One popular tool combining these steps is Herbie~\cite{herbie}.
Herbie uses sampling techniques to identify floating-point error;
  constructs candidate rewrites using algebraic and analytic identities,
  and tests those rewrites against higher-precision executions
  to identify the rewrite with the lowest floating-point error.
In recent releases, Herbie can output multiple suggestions
  with different performance and accuracy characteristics~\cite{pherbie}.

Unfortunately, all of these tools, Herbie included,
  are difficult for developers to use and integrate into their workflows.
Users are typically expected to
  identify the expression and inputs of interest up front;
  compare them to other sources or the user's own ideas;
  and make trade-offs between accuracy and other goals
  (e.g., maintainability),
  all without tool support.
Users are often recommended
  to switch between their code editor, version control system,
  a mathematical visualization tool, and multiple Herbie instances
  in order to solve a single problem~\cite{kneusel-numbers}.
VSCode-PRECiSA \cite{vscode-precisa}, 
  a VSCode interface for the PRECiSA 
  command-line tool \cite{precisa} designed to support the process 
  of analyzing a single program in several ways,
  is somewhat of an exception; 
  however, it does not address the problem of tool
  interoperation.
We developed Odyssey to address these limitations
  by providing a single integrated workbench
  for the full floating-point rounding error workflow.
To lower the barriers to adoption,
  Odyssey uses Herbie, a widely used and open source
  tool~\cite{herbieGithub,kneusel-numbers},
  under the hood.

\subsection{Expert tools for design space search}
Odyssey is an expert tool for numerical analysts to re-write floating point
expressions. Unlike tools for end-users, expert tools are designed for users
with extensive design and implementation experience. Experts have honed
specialized workflows, leverage insights to improve upon automated or
semi-automated approaches, and are comfortable wading into low-level details.
For example, expert developers optimize the performance of
applications~\cite{cito2018performanceHat} and specialized pipelines. In the
domain of high-performance image processing,
Roly-Poly~\cite{ikarashi2021guidedOptimization} is a system built on top of the
Halide compiler~\cite{ragan2013halide} for expert engineers to explore
trade-offs and decide among possible optimizations. Odyssey is similar to
Roly-Poly in that it supports interactive workflows with an automated tool to
support expert users. In the statistical analysis domain, multiverse analysis
tools such as Boba~\cite{liu2020boba} and Multiverse
Debugger~\cite{gu2023understanding} enable expert statistical analysts to assess
the robustness and sensitivity of analysis results. The intended users are
experts in statistical analysis but not necessarily in multiverse authoring.
Similarly, Tsugite helps expert fabrication users create new wood
joints~\cite{larsson2020tsugite}. Odyssey adds to this growing body of research
on expert tools for a new domain, and we discuss key insights that 
could serve as design principles generalizable across domains (\autoref{sec:discussion}).


\begin{figure}
  \centering
  \shadowimage[width=\linewidth]{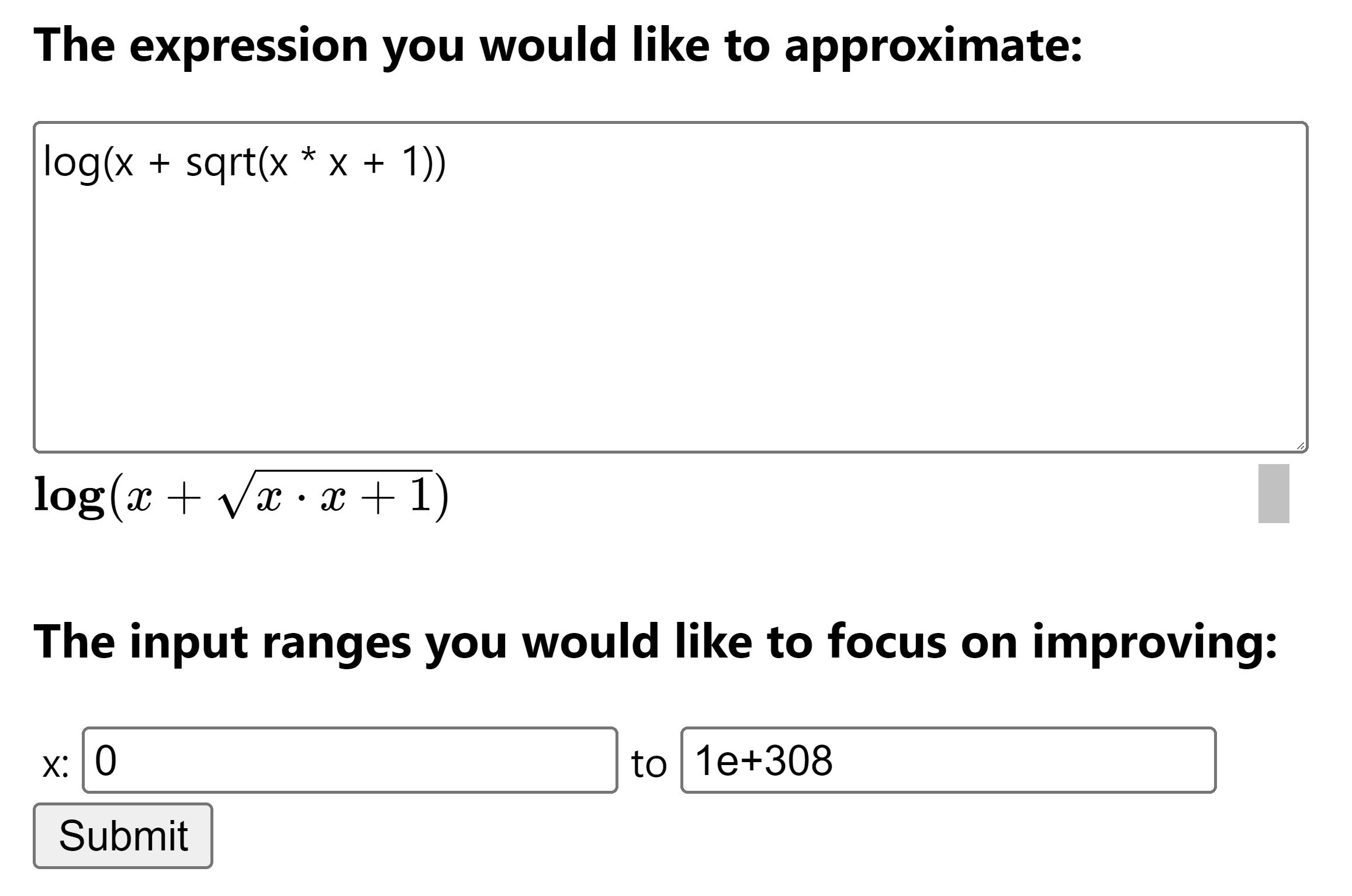}
  \caption{Users enter a new expression.}
  \label{fig:new-tab}
\end{figure}

\begin{figure*}
  \centering
  \includegraphics[width=\linewidth]{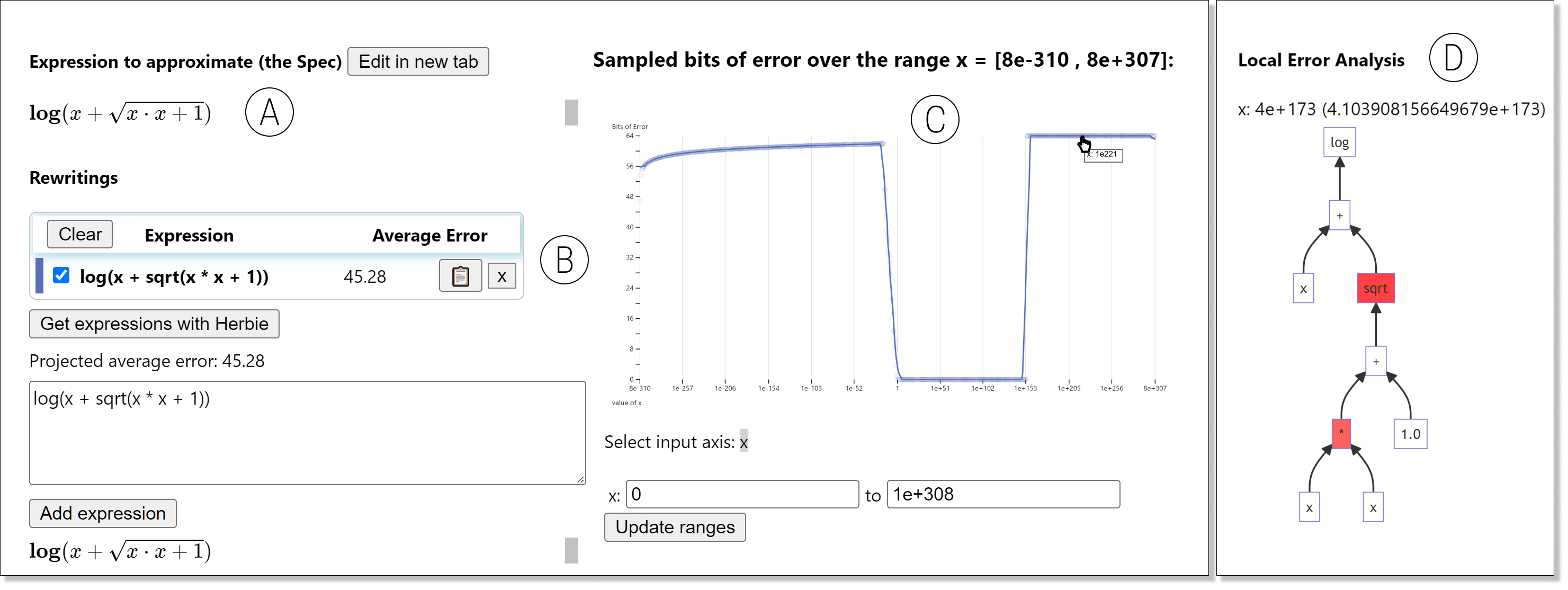}
  \caption{Diagnosis.
  The specification (A) shows the expression the user is trying to implement.
  The rewritings table (B) shows the expressions the user has tried.
  The error plot (C) shows the error of the current expression.
  The local error heatmap graph (D)
    shows the error breakdown of the currently selected point.
  }
  \label{fig:diagnosis}
\end{figure*}

\begin{figure}%
  \centering
  \shadowimage[width=.8\linewidth]{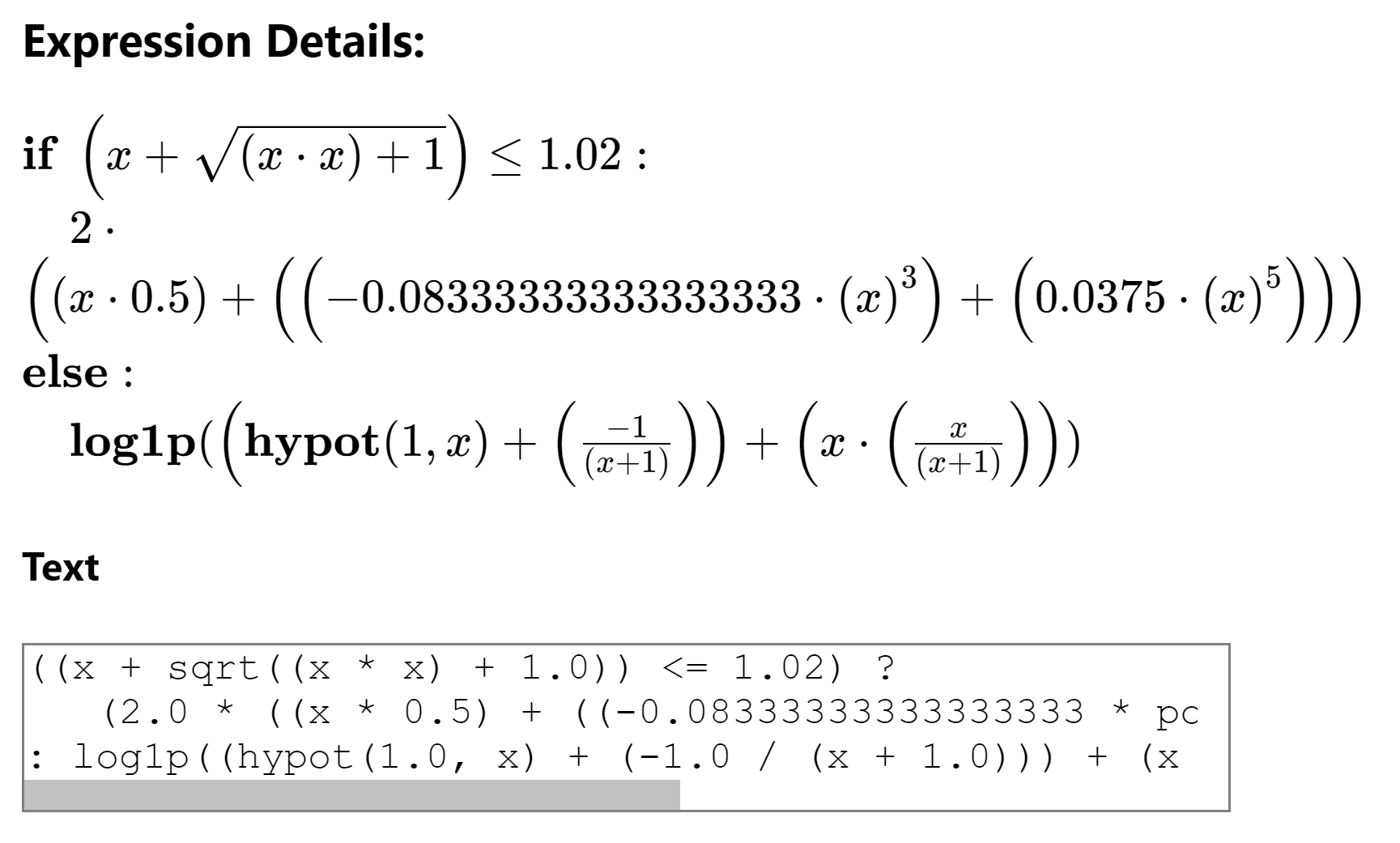}
  \caption{The Expression Details view shows 
  a LaTeX rendering and plain text to help users 
  understand and work with the selected expression.
  }
  \label{fig:expression-details}
\end{figure}

%


\section{Usage Scenario}

Alex, a numerical analysis expert, 
  has received a report that there is an issue in 
  the \texttt{asinh} function of a popular programming language's standard library.\footnote{As mentioned earlier, this issue is based on a
  real-world problem that a numerics expert recently found and addressed for
  the Rust standard library using Herbie~\cite{herbie-rust}}

Alex now needs to develop
  an accurate implementation of the \texttt{asinh} function.

\subsection{A Typical Debugging Process}
The \texttt{asinh} function is defined, for positive $x$, as
  $\operatorname{asinh}(x) = \log(x + \sqrt{x^2 + 1})$.
Based on the report, 
  Alex hypothesizes that the issue involves
  the high range of the function's input.
The $x^2$ term will overflow for large $x$.

They aren't immediately sure how to fix this.
They turn to a state-of-the-art automated tool,
  Herbie, for help.
Alex runs Herbie on the \textit{asinh} expression.
Herbie suggests a replacement expression
  and shows an error plot 
  for the original and final expressions.
  

Alex wants to start rewriting, but now faces a series of obstacles.


First, 
  Herbie's error plot suggests that 
  there is another source of error
  in the expression---small inputs, between 0 and 1.
Alex needs to \textit{diagnose the cause} of this error by finding a subexpression to rewrite.
Alex sets up a REPL for the math library 
  and manually steps through each subexpression.
Alex considers
  its input and output ranges 
  to see where errors occur.

Second, 
  Alex needs to \textit{generate new solutions} and test them.
Although Herbie suggests a potential rewrite,
  it is still error-prone for small inputs.
Drawing on their experience, 
  Alex wants to try out new expressions, 
  but Herbie does not support this.
As a result, Alex abandons Herbie, 
  writes a new expression, 
  and sets up a new testing framework.
Alex is frustrated that 
  they have to figure out how to set this testing up by themself,
  even though Herbie has internal tools that are capable of this.
Future iterations will require Alex to start all over, 
  discouraging them from exploring and 
  finding an expression with more desirable error characteristics. 


Third,
  Alex finds two rewrites which fix adjacent parts of the domain,
  and now wants to join them.
This requires \textit{tuning} the constant used for picking the branching point.
However, in Alex's current test framework,
  the consequences of changing the constant are not evident.
In other words, an iterative design process is not supported.


In order to address the above issues,
  Alex spends hours stitching together workarounds.

Alex needs an integrated tool 
  designed for human-directed expression debugging and interactive rewriting.
Odyssey is designed to help experts like Alex who fix floating-point bugs that impact the core of a programming language.

\subsection{Using Odyssey}
\subsubsection*{First Stage: Diagnosing Problems}

Using Odyssey, Alex begins by typing the mathematical definition,
  \texttt{log(x + sqrt(x * x + 1))},
  into Odyssey's expression entry box (see~\Cref{fig:new-tab}),
  along with the range of possible \texttt{x} values.
In this case, the definition is only valid for positive $x$,
  so Alex enters $0$ as the lower bound.
  Since this is a library function that can be executed on any input,
  Alex leaves the default upper bound of $10^{308}$ in place.
The expression and initial input range
  are used to initialize Odyssey's main screen (\Cref{fig:diagnosis})
  and appear in the top left corner of the screen (\Cref{fig:diagnosis}A).
If the user needs to launch multiple Odyssey sessions,
  this part of the screen will help them differentiate them.

Beneath the initial expression, Alex sees Odyssey's rewritings table~(\Cref{fig:diagnosis}B).
The rewritings table allows the user to collect
  multiple versions (or ``rewritings'') of the expression
  and compare them for accuracy.
Each rewriting in the table shows its average accuracy,
  and rewritings can also be selected or hidden
  to control the display of other information in Odyssey.
Initially, the rewritings table contains a single rewriting,
  the direct implementation of their expression.
In this example, the initial rewriting
  has quite high error (45.28 bits
   out of 64)
  indicating that there is quite some work left to do
  to produce an accurate implementation.
  
To better understand the source of this error,
  Alex refers to the error plot (\Cref{fig:diagnosis}C).
This plot shows the error of every rewriting in the table.
The horizontal axis shows different input values $x$
  spanning hundreds of orders of magnitude;
  the vertical axis shows error, with higher values being worse.
In this example, three regions are clearly visible:
  inputs $x < 1$, with high error;
  inputs $1 < x < 10^{150}$, with low error;
  and inputs $10^{150} < x$, with high error again.
Distinct regions like these often have distinct causes of error
  and are a starting point for exploring more deeply.

To begin investigating, Alex clicks on one of the points in the error plot;
  this updates Herbie's ``local error heatmap'' display (\Cref{fig:diagnosis}D).
Local error is an internal heuristic in Herbie
  that identifies which operations in a rewriting
  cause rounding error at a given point.
By clicking on one point with $10^{150} < x$,
  and another point with $x < 1$,
  Alex confirms that this expression
  has two distinct sources of error:
  for large inputs $x$, the source of error is 
  the \texttt{sqrt} and \texttt{*} operations,
  while for small inputs $x$, the source of error
  is the \texttt{log} operation.
After diagnosing the operations with error 
  and the affected inputs,
  Alex begins generating solutions
  to these floating-point rounding error problems.

\begin{figure*}
  \centering
  \shadowimage[width=\linewidth]{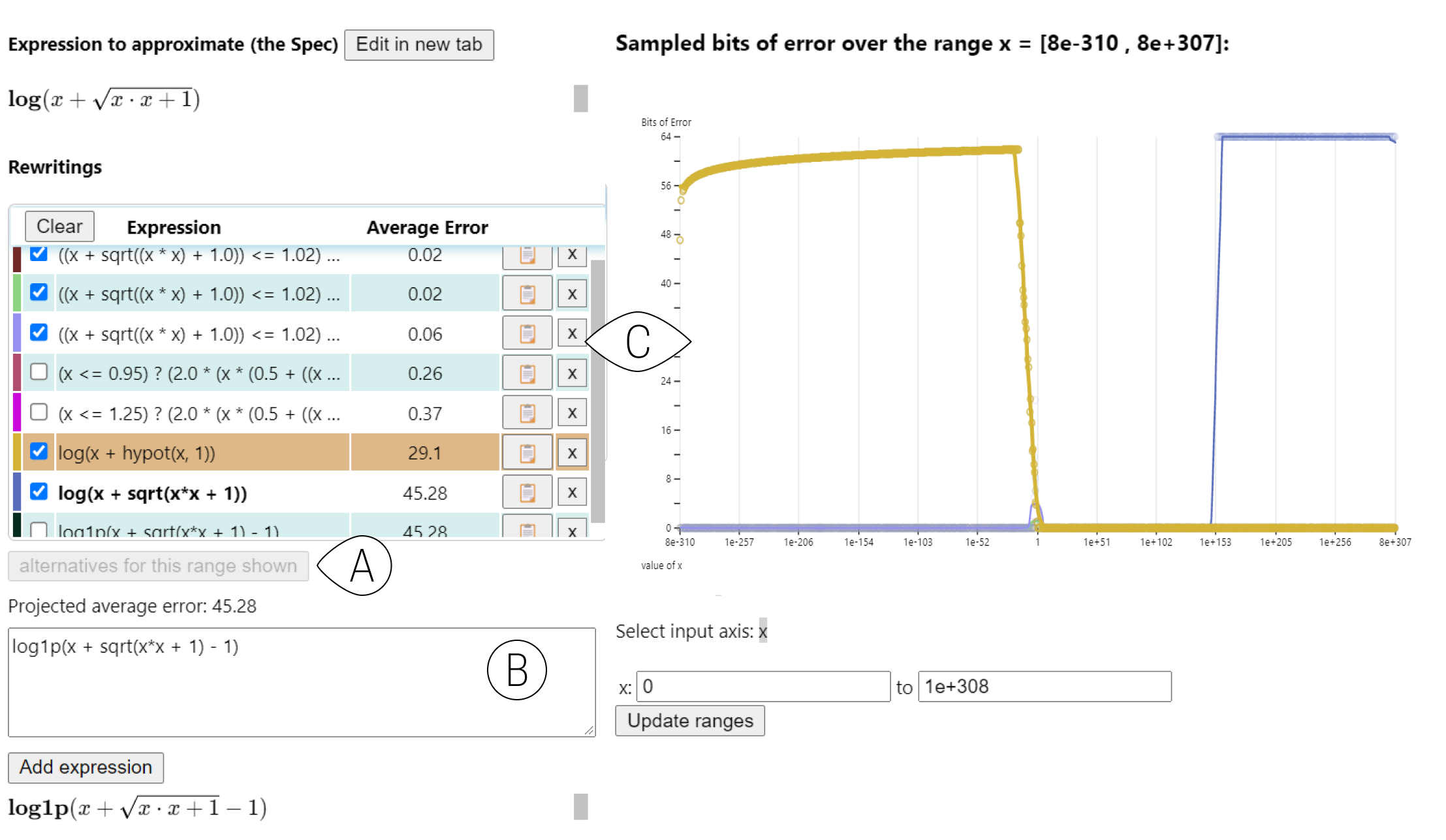}
  \caption{Solution generation. 
  User can request rewritings from Herbie by pressing a button (A)
    or enter their own using the expression edit box (B), which provides
    live feedback and estimates the expression error on the current sample.
  The rewritings table and the error plot (C) are updated every time a rewriting
    is added, allowing the user to compare the quality of different rewritings.}

  \label{fig:generation}
\end{figure*}

\subsubsection*{Second Stage: Generating Solutions}

To start generating solutions quickly,
  Alex queries an automated tool using
  the ``Get expressions with Herbie'' button
  (\Cref{fig:generation}A).
This automatically translates the expression
  into Herbie's input format;
  invokes Herbie;
  evaluates the error of each of Herbie's suggestions;
  and translates each one back to a human-readable format.

In this case, invoking Herbie
  adds five suggestions to the rewritings table
  and to the error plot (\Cref{fig:generation}C).
Since each rewriting in the table lists its error,
  Alex sees immediately that Herbie's suggestions
  reduce the original 45.28 bits of error
  to as low as 0.02 bits of error.
Moreover, each rewriting's error is also graphed on the error plot,
  with different rewritings shown in different colors.
Users can highlight the plot for an expression 
  by clicking on its row in the table.
For example, by clicking Herbie's fifth suggestion,
  \texttt{log(x + hypot(1, x))},
  Alex sees that this expression avoids
  error for $10^{150} < x$
  but still has error for smaller values of $x < 1$.
Multiple suggestions will probably need to be consulted,
  compared, and combined
  to achieve Alex's accuracy, performance, and maintainability goals.

Herbie is not the only source of rewritings in Odyssey.
In fact, human creativity is often needed
  to overcome roadblocks for automated tools,
  and rewritings may also be sourced from other tools,
  from papers, or from online references.
Therefore, Odyssey allows Alex to add rewritings
  directly to the rewritings table using the edit box (\Cref{fig:generation}B).
As they type,
  their expression is automatically rendered
  and an error estimate is provided,
  to help avoid typos and other low-level mistakes.
As Alex works on this expression,
  the table of rewritings will grow to contain
  all of the various rewritings or ideas they have considered.
By leaving this basic organizational task to Odyssey,
  Alex is able to focus on high-level reasoning.

\subsubsection*{Third Stage: Tuning}

\begin{figure*}
  \centering
  \includegraphics[width=\linewidth]{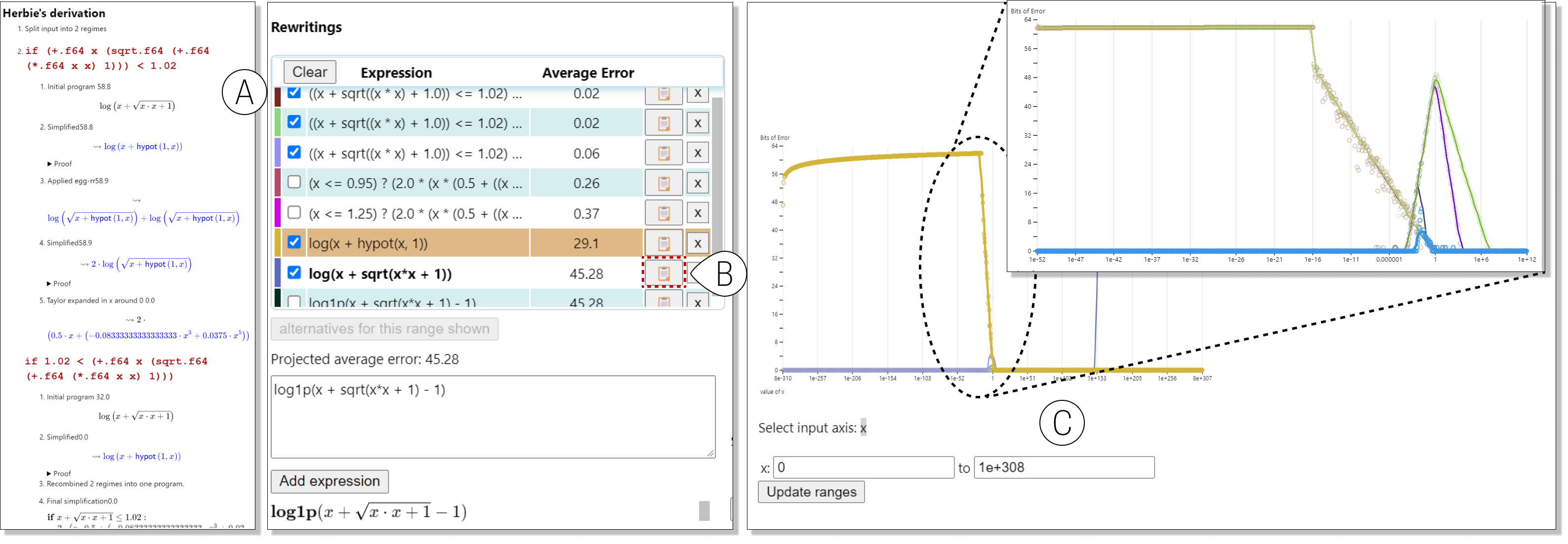}
  \caption{Tuning. 
  The user can use derivations (A) to help them understand Herbie-generated rewritings.
  Each expression can be copied using the copy button (B) for easy editing of
    existing rewritings.
  The user can use the input range editor (C) to ``zoom in'' on critical ranges---%
    i.e., resample and reanalyze all expressions on a new range. 
    Above, the user has tried rounding some of an expression's constants 
    after zooming.
  }
  \label{fig:tuning}
\end{figure*}

After generating solutions
  to the various floating-point issues in this expression,
  Alex wants to understand how these rewritings can be combined
  to produce a single implementation of the expression
  that satisfies their accuracy, performance, and maintainability goals
  (\Cref{fig:tuning}).

Since, in this case, many of the rewritings are generated by Herbie,
  they start by understanding those rewritings
  in greater depth.
To do so, Alex clicks on one of these rewritings
  and looks at the derivation provided for it (\Cref{fig:tuning}A).
The derivation of a Herbie-generated rewriting
  shows the sequence of steps Herbie used to produce it.
Alex scans one derivation that has caught their attention
  both for ideas
  that can be lifted and combined with a different rewriting,
  as well as for potentially dangerous steps.
In this case, they spot
  that Herbie used a Taylor series expansion
  to derive one of the rewritings.
Taylor series expansions are dangerous,
  because they are often valid only for inputs in a certain range,
  and can lead to high error if used outside of that range.
In this case, Herbie guarded the Taylor series
  with the conditional $x \le 1$;
  however, it may be possible to tune the condition further.

To begin tuning this piece, Alex uses
  Odyssey's range adjustment control (\Cref{fig:tuning}C).
Since the conditional has a threshold at $1$, 
  Alex enters a range of inputs near $1$: 
  $10^{-52} \le x \le 10^{12}$.
When Alex updates the range,
  Odyssey samples a new set of inputs
  all chosen from the selected range,
  and the plot updates to show only the new set of inputs.
Because these inputs are all clustered near $1$,
  Alex can now examine error in this range
  at much higher resolution.
Here, the higher resolution reveals
  what inputs around 1 have a spike in error.

To fix this new-found problem,
  Alex continues to test new rewritings
  using the expression edit box.
Since, at this point, Alex has already found
  many quite-accurate rewritings,
  they choose to modify an existing rewriting
  using the copy-to-clipboard button (\Cref{fig:tuning}B).
This allows Alex to easily make small adjustments,
  such as raising or lowering the threshold 
  by rewriting the branch condition,
  and see how that affects the inputs they have focused on.
Alex may not always tune expressions for accuracy;
  they might instead simplify rewritings to make them run faster,
  or make modifications to improve readability and maintainability.
In those cases, the error graph allows Alex to validate
  that error has not increased unacceptably.
Finally, Alex has tuned the expression to their liking,
  so they use the copy-to-clipboard button
  to copy the final expression 
  and insert it into their program.

Reviewing these steps,
  Alex
  used a three-step floating-point error improvement workflow:
  diagnosing the sources of floating-point error;
  generating candidate solutions to these source of error;
  and then tuning and validating the resulting solution
  until it met their accuracy, performance, and maintainability goals.
The entire process was orchestrated
  through Odyssey's table of rewritings and error plot,
  which track the various rewritings Alex already considered
  and allow Alex to easily compare rewritings over the input range.
Odyssey additionally provided convenient ways
  to leverage the automated error-improvement tool Herbie,
  including invoking Herbie, visualizing internal heuristics,
  and presenting derivations.
Combined, these features allow Alex
  to focus on higher-level concerns such as accuracy-improving rewrites
  and acceptable trade-offs between their goals.

\section{Iterative design process}

To understand how to meet the needs of Herbie's users,
  after reviewing user-submitted bug reports,
  testing changes to the existing Herbie user interface,
  and mocking up a new interface,
  we conducted an iterative user design study.
As we observed users working with the prototype,
  we identified new needs and added features
  to meet those needs.

\subsection*{User Design Study with Prototype}

Our user design study consisted of nine interviews
  with participants ranging from floating-point novices to experts.
Most participants were graduate students
  working on floating-point-related research
  with at least two years of experience.
We spaced these interviews out
  and iteratively added features to Odyssey,
  responding to user concerns after each interview.
We made the following observations.


First, we found that more experienced users
  iteratively submitted many hand-written programs to Odyssey.
In some cases, users modified a Herbie result,
  used Odyssey's reported error to confirm
  that the change didn't harm accuracy,
  and then used the modified program
  as a base for further modifications.
In other cases, users modified a Herbie result
  and re-ran Herbie on the modified expression,
  helping Herbie around a road-block of some kind
  and achieving a lower error as a result.
We also saw users combining pieces of different programs
  into a single final program.
Users described implicit trade-offs,
  for example noting that Herbie's result was very complex,
  and that deleting certain terms from Herbie's result
  was less accurate but easier to read.

\begin{figure*}
  \centering
  \includegraphics[width=\linewidth]{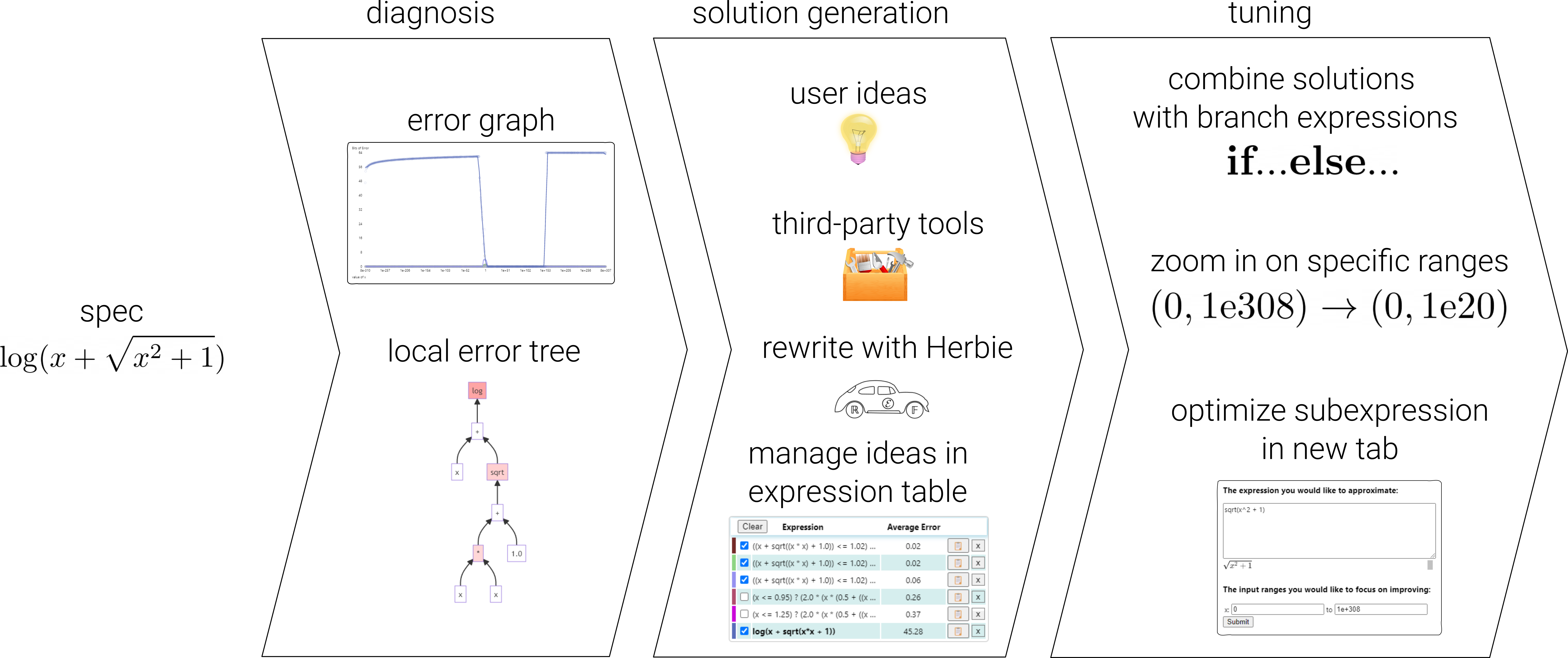}
  \caption{The general workflow supported by Odyssey.
  Odyssey starts with a real-number specification,
    analyzes sources of error,
    creates different solutions based on the analysis,
    and tunes solutions based on user's needs.
  }%
  \label{fig:workflow}
\end{figure*}

Second, we noticed that many participants,
  including both novices and experts, 
  struggled to explain \textit{why} there was error in an expression,
  even when they could see the error in Odyssey's error plot.
For example, in the program $\log(x + \sqrt{x^2 + 1})$,
  most users could guess
  that the error for large $x$ values was caused by overflow, but
  far fewer participants could identify that error for small $x$
  was caused by the $\log()$ operation.

  In a follow-up conversation with the Herbie developers,
  we learned that Herbie used a metric called ``local error''
  to identify which operations were likely sources of error.
We decided that exposing this metric to the user
  as a local error ``heatmap'' (see~\Cref{fig:diagnosis}D)
  could help users better understand floating-point error.
Participants immediately began using per-point local error 
  to explain why error occurred for specific inputs to specific programs.
In this process, we also discovered
  that Herbie's local error implementation
  had a subtle bug on specific, rare inputs, leading to a patch.


Finally, after initially removing derivations (see~\Cref{fig:tuning}A),
  we realized that they were an important foundation for users' trust
  in Herbie's results.
For example, one participant was surprised
  when Herbie recommended the expression ``1.0''
  as an ``improved'' version of some much more complex expression
  and became skeptical of all of Herbie's other outputs,
  manually performing derivations to check
  that those expression had been computed correctly.
Adding back support for derivations
  gave users more trust in Herbie's suggestions.

Through the user design study, we observed the following:
\begin{itemize}[label=$\circ$]
  \item Experienced users follow an iterative process when rewriting expressions.
  \item Rapid feedback during expression input
    helps users catch low-level mistakes.
  \item Users need help understanding what part of the expression is causing error.
  \item Users want justification and explanation for the steps of automated tools.
\end{itemize}

\section{Expression Rewriting Workflow and Design objectives}

Our design study 
  led us to model floating-point error improvement
  as a well-defined workflow consisting of three main stages:
  diagnosis, solution generation, and tuning.


\subsubsection*{First Stage: Diagnosing Problems}

In this stage, users identify problematic operations within expressions,
  determine which problems are relevant to their objectives,
  and finding starting points for further analysis.
For instance, in an expression like $\log(x + \sqrt{x^2 + 1})$,
  users must determine that
  the $x^2$ operation overflows for large values of $x$,
  while the logarithm is inaccurate for small values of $x$.
The user then decides whether
  large values of $x$ are relevant in their environment.
  If so, they focus on avoiding the overflow in $x^2$.

We developed two principles to support diagnosis.
First, users need ways to focus analysis
  on the parts of the input range and expression
  they care about investigating---%
  without losing track of the broader analysis.
Second, even experts 
  need tools to help determine which operations cause error
  without relying on their expertise
  or resorting to trial-and-error operation replacement.

\subsubsection*{Second Stage: Generating Solutions}

In the second stage,
  users gather potential rewritings from a variety of sources.
The objective is to create a pool of rewritings
  that the user can evaluate and combine
  to address the problems identified in the first stage.
While existing tools, like Herbie,
  are a valuable source of ideas and potential rewritings,
  the user must still track and organize the outputs.
Moreover, rewriting ideas may come from many other places:
  other automated tools, papers, online references,
  and even the user's own creativity.
Users need to collect the available rewritings,
  keep track of their origin,
  and organize them for easy evaluation.

We developed three principles to support solution generation.
First, there must be a central repository
  of rewritings drawn from multiple sources.
The repository must also store source-specific details,
  such as Herbie's derivations.
Second, since users themselves are a major source of ideas,
  manual input of rewritings must be supported,
  with instantaneous feedback to provide low-level error checking.
This supports a tight feedback loop and eases iterative exploration.
Third, where possible, it should be possible to use user inputs
  as starting points for additional automated exploration,
  allowing users to overcome roadblocks faced by automated tools.

\subsubsection*{Third Stage: Tuning}

In the third stage, users test, compare, and tweak rewritings
  to optimize for their accuracy, performance, and maintainability goals.
Often the diagnosis and solution generation phases help users identify multiple independent problems
  and multiple independent rewritings that address them.
Users must combine these rewritings to address error.
This combination process is itself iterative.
  Users needing to validate
  that the combination did not introduce its own error.
Moreover, the combination process might itself need tuning.
  Users may want to adjust the threshold at which
  they switch from one rewriting to another.
Overall, this stage involves iterative refinement and experimentation 
  until the user is satisfied with the result.

We developed two principles to support tuning.
First, the user needs ways to compare rewritings for accuracy
  and get instantaneous feedback as they work.
Second, users need explicit support for combining rewritings,
  whether directly using ``if'' conditions
  or indirectly by allowing the user to see multiple rewritings at once.

The order of these stages is not fixed, and users may iterate between them,
  but we think these principles address explicit user
  needs during floating-point error improvement
  with an automated tool.



\section{Implementation}

Odyssey is implemented in two pieces:
  a ``backend'' that uses Herbie to dispatch numerical tasks,
  and a ``frontend'' implemented using web technologies
  to present an interactive workbench UI to the user.
Odyssey can be used via a web browser
  or embedded into tools like
  Visual Studio Code.

\subsection{``Database Workbench'' Architecture}

The key to supporting our design principles
  is Odyssey's ``database workbench'' architecture.
In this architecture,
  Odyssey stores a list of rewritings that the user
  is exploring and makes calls to independent analysis, visualization, 
  and generation tools that run on the backend.
This architecture stores all of the state on the frontend,
  allowing direct manipulation by the user.
The automated analysis, visualization, and generation tools,
  meanwhile, are stateless,
  being invoked by Odyssey on whatever rewritings
  the user is currently considering.
This architecture puts the user at the center at the center of the search.

This architecture also leads to a natural separation of concerns
  between the frontend and backend.
The Odyssey frontend implements all interactions,
  graphics, and manipulation actions.
However, all numerical tasks
  (sampling, evaluating error, and generating expressions)
  are the responsibility of the backend.
This 
  ensures proper support for 
  low-level operations
  like enumerating floating-point numbers
  and other numerical tasks that
  depend on the user's target environment.
While Odyssey currently only invokes Herbie subsystems,
  the backend is intended to invoke other tools as well.
  
\subsection{The Odyssey Frontend}

The Odyssey frontend provides a rewritings table and error plot
  to help users diagnose problems, generate solutions, and tune the results.

The main state is stored in the rewritings table,
  shown in \Cref{fig:generation}.
All rewritings the user is considering---%
  including both those generated by Herbie
  and those entered by the user,
  are stored here.
Each rewriting also shows its average error,
  for easy comparison.
A checkbox allows the user to hide expressions
  from the error plot and other parts of the UI,
  which functions as a kind of ``archiving'' operation
  so that users can ignore sub-par rewritings
  without an irreversible deletion operation.
Additionally, a clipboard button
  allows users to copy rewritings,
  which is essential to users modifying or combining rewritings.
None of these interactions involve the backend,
  and are thus instantly responsive to user action.

The input box allows adding rewritings to the table
  using a natural mathematical syntax
  backed by a parser from the mathjs library~\cite{mathjs}.
Odyssey then converts that input
  both to an instantly-updating LaTeX render
  (to help users catch mistakes and typos)
  and to the standard FPCore input format,
  which Herbie uses to represent rewritings.
Herbie is then invoked to analyze the error of the new rewriting,
  which is then added to the plot.
Additionally, rewritings can be added to the table
  by invoking Herbie to generate suggested rewritings;
  any rewritings suggested by Herbie are also
  converted from FPCore back to LaTeX and mathematical syntax
  so that the user does not have to understand FPCore
  in order to use Odyssey.

The main visualization is a large error plot.
This plot shows the error on all of the sampled inputs,
  for each of the rewritings in the rewritings table,
  with colors helping users match each rewriting to its error plot.
Because rewritings often have identical error over some range
  the user can click on a rewriting in the table
  to highlight it in the error plot;
  users can also use checkboxes in the table
  to hide expressions from the error plot.
By hovering over each point in the error plot,
  the user can see the exact sampled input,
  and by clicking on a point,
  they can update parts of the UI
  (such as the local error heatmap)
  to focus on that specific input.
The user can also adjust the input domain
  using an input range selector below the plot.
Changing the input domain causes Odyssey
  to resample inputs, evaluate each rewriting on the new inputs,
  and redraw the error plot using the newly-evaluated errors.
Once again, besides adjusting the input range,
  all operations are instantaneous and do not invoke the backend.

On its own, Odyssey does not provide any additional features.
However, Odyssey is extensible,
  and tools invoked by the backend
  can offer additional visualizations.
To see these additional visualizations,
  the user selects a specific rewrite,
  and the visualizations are shown beneath the main UI.
Selecting the specific rewriting
  means that different rewritings,
  which might come from different sources,
  can provide different kinds of justifications or explanations.
Our Herbie backend provides two such visualizations:
  the local error heatmap and derivations.
When Odyssey is extended to support additional backend tools,
  we expect each tool to provide its own additional visualizations.

\subsection{The Herbie Backend}

Odyssey's Herbie backend is used
  to sample inputs, evaluate the error of rewritings,
  and suggest new rewritings to the user.
Herbie was originally designed as a batch-mode tool,
  so part of our work involved adding an HTTP API 
  to expose various internal analysis functions
  so that they can be invoked by Odyssey.
Luckily, the Herbie features that we wanted to expose,
  including input sampling and error evaluation,
  were already independently-invocable functions in Herbie.

A key challenge in the backend is dealing with latency.
Herbie's initial design as a batch-mode tool 
  means that Herbie typically
  samples inputs, evaluates error, and suggests rewritings
  every time it is invoked,
  even though some of those steps (like sampling inputs) are slow
  while others (like evaluating error) are fast.
To address this, Odyssey's Herbie backend
  independently caches the outputs of each step (like the sampled inputs).
This way, evaluating the error of an expression
  is done on cached sampled inputs and takes milliseconds
  instead of resampling the inputs, which would take seconds.

Further, all of Odyssey's invocations of the backend are asynchronous,
  allowing the user to continue working
  while Herbie processes their requests.


By keeping the latency of most operations under a second
  and offering access to previously-inaccessible heuristics 
  like local error (an internal search heuristic) 
  and expression derivations (previously a debugging tool for Herbie developers),
  Odyssey's ``database workbench'' architecture
  allows users to stay in the flow of their work as they solve rewriting problems.


\begin{table}
  \centering
  \begin{tabular}{|c|p{6cm}|}
\hline
  Expert \# &  Background: \\
\hline
    1 &  Industry, FP hardware + supercomputing (number systems for minimization problems), 45+ years. \\
\hline
    2 &  Professor, FP tools (mixed-precision conversions and program analysis), 10+ years. \\
\hline
    3 &  Grad student, FP hardware (datapath optimization), 5 years. 
    \\
\hline
    4 &  Professor, verification (correctness analysis), 9 years. \\
\hline
    5 &  Industry, FP hardware (interval analysis, transcendental functions), 3 years. 
    \\
\hline
\end{tabular}

  \caption {Five experts from the floating-point community evaluated and suggested future directions for our work.}
  \label{fig:experts}
\end{table}

\section{Expert Evaluation}
The goal of the expert evaluation
  was to assess the effectiveness of Odyssey
  in supporting the three-stage workflow we identified:
  diagnosing problems,
  generating solutions,
  and tuning expressions.

\subsection{Protocol}
We conducted an interview study
  with five experts from the floating-point community
  (see \autoref{fig:experts})
  to evaluate the effectiveness of Odyssey
  in supporting the three-stage workflow.
We recruited the experts via email
  through professional networks and communities (e.g., FPBench).
Each expert had different levels of experience
  in academia and industry,
  ranging from 3 years to over 45 years,
  and their backgrounds covered various aspects of floating-point systems,
  including hardware design, verification, and optimization.

Each interview session was conducted over Zoom, 
  with experts operating the tool via remote control 
  to avoid early issues we experienced with participants on networks
  with special configurations.
Interviews lasted between 60 and 90 minutes
  and consisted of three parts:
\begin{itemize}
  \item \textit{Introduction and tutorial.} 
The first author briefly introduced Odyssey 
  and the problems it is designed to address. 
Then, each expert followed a hands-on tutorial 
  demonstrating the usage of Odyssey on a simple example.
\item \textit{Seven tasks.} Each expert completed seven tasks, each designed
for one of the three workflow stages and aimed at eliciting the experts'
reactions to different parts of Odyssey's interface (\autoref{fig:tasks}). If the experts
encountered difficulties, the first author provided guidance or reminded
them of relevant interface features from the tutorial.
\item \textit{Exit survey and discussion.} To conclude, each expert
completed a survey (\autoref{fig:survey-results}) 
and participated in a semi-structured interview with the
first author, where the experts reflected on their experience with Odyssey
and provided feedback on potential improvements and extensions. The first
author specifically asked for experts' opinions on the legitimacy of the
workflow we aim to support, its relevance to their work, and the extent to
which they felt Odyssey supported each part of the workflow.
\end{itemize}
Throughout all three parts, the experts' screens and audio were recorded. The
first author also took note of the experts' comments, insights, and responses to
the tasks and survey questions. All study materials are provided as supplemental material.

\subsection{Analysis and Results}
We conducted an iterative, thematic analysis of expert solutions and the first
author's notes for each stage of the workflow. Below, we discuss the experts'
responses to the relevant tasks and survey items for each stage. Through this
analysis, we aim to provide a qualitative evaluation of Odyssey's effectiveness
in supporting each part of the workflow.

\begin{table*}
  \centering
  \begin{tabular}{|c|p{5cm}|p{6cm}|c|}
  \hline
      Task & Description & Targeted part of workflow & Success rate \\
  \hline
  1 & $log(x + \sqrt{x \cdot x + 1})$ is an expression for the inverse hyperbolic 
  sine. Identify the parts of the expression causing errors for large/small $x$. & Diagnose troublesome 
  subexpressions and problematic ranges. & 4/5 \\
  \hline
  2 & Use Odyssey to find a solution for the troublesome square root subexpression from task 1. 
  & Generate solutions for the subexpression and use these to optimize original expression.
  & 4/4 \\
  \hline
  3 & Is your solution to task 2 good enough? 
  & Use visualizations to form evaluation criteria for ending analysis.
  & 4/4 \\
  \hline
  4 & Identify problems with branch expressions in fully automated solutions for task 1.
  & Explain important features of expressions and diagnose issues. & 3/4\\
  \hline
  5 & Use Odyssey to find and recommend \textit{log1p} to solve small $x$.
  & Nudge an automated tool past roadblocks to generate better solutions.
  & 2/3 \\
  \hline
  6 & Evaluate whether the full solution for the expression after tasks 1-5 is trustworthy.
  & Use Odyssey's feedback on expressions and information about expression soundness to 
  evaluate expressions' trustworthiness and fitness based on personal standards.
  & 2/2 \\
  \hline
  7 & Use branch conditions to outperform a fully automated rewriting for the expression
  $(exp(x)-2) + exp(-x)$. & Mix solutions from different sources and tune branch conditions
  to create stronger solutions. 
  & 3/4 \\
  \hline
\end{tabular}

  \caption {Experts worked through up to seven tasks to exercise the features of Odyssey before a survey-based discussion. Due to time constraints, 
  not all experts completed all tasks.}
  \label{fig:tasks}
\end{table*}

\subsubsection*{First Stage: Diagnosing Problems}

Task 1 required experts to analyze the error in an inverse hyperbolic sine
implementation and identify the parts of the expression causing errors, then
decide which operation or operations needed to be rewritten in order for the
rewriting to correctly handle large inputs. Among the five experts, four
successfully completed this task, relying on Odyssey's error visualizations
(see~\autoref{fig:diagnosis}).

P2 explored multiple input ranges in order to identify 
  the two problematic operations: 
  \longquote{This is across the
entire sample... so I wonder if it's doing something different on this side
[clicking a point with a small $x$ value and looking at the local error graph] So
there it's all the log, and over there... [clicks a large $x$ value] it's all the
square root. So that's interesting, it's actually coming from different
operations.} Here, the error plot effectively surfaced the two areas of high
error (small and large $x$ values), giving the expert clear places to look for
troublesome operations. Then, by switching between inputs in different regions,
the expert was able to see that the problematic operation was different
between these regions.

In the survey (see item 3 in~\autoref{fig:survey-results}),
  all five experts rated the interface's ability
  to help identify or confirm specific problems with expressions 
  at a 7 out of 7.
We attribute this success mainly to the error plot and local error heatmap,
  which implemented the second principle we identified for a good diagnosis tool.
  They supported the user in assigning responsibility for error
  without relying on expertise or resorting to trial and error.
As P2 concluded, \longquote{%
  Having the graph and being able to click 
  on the different places where error is high 
  is definitely nicer than just looking at output 
  in a text file.}

\subsubsection*{Second Stage: Generating Solutions}

We designed several tasks to evaluate Odyssey's support for collecting and
evaluating new expressions that address the identified problems in floating-point
expressions. Close to all experts who attempted each
task succeeded (see Tasks 2 and 5 in~\autoref{fig:tasks}).

Task 2 required experts to analyze a troublesome
subexpression from Task 1 and find a better rewriting for it. Then, experts needed
to bring the solution back to the original analysis and decide if they were
happy with it. Four of the five experts who attempted Task 2 successfully
completed it, showing that the interface facilitated the collection of solutions
and their integration into existing expressions. Of those four, two experts found their own unique approaches to solving the problem identified in Task 1 rather than relying on an automated solution. One expert pulled a factor of $x$ out of the square root, and another expert created a branch that switched to an approximation for large values of $x$. Both of these approaches showed low error on the error plot, though the experts noted there could be issues with these choices (for example, branching impacts performance, and dividing by $x$ is risky when $x$ could be 0). This showcases the flexibility of
Odyssey in allowing users to explore alternative solutions and evaluate their
impact on the error plot. 
(The expert who did not complete Task 2 was our first participant,
with whom we lost much of the interview time due to the networking issues mentioned earlier.)

Similarly, Task 5 asked experts to find a more accurate rewriting for a
subexpression applicable to small values of $x$. Three out of the four
experts successfully completed this task, further supporting the
effectiveness of Odyssey in assisting experts in gathering and evaluating
potential solutions.

P3 had the following to say about working through the process up to Task 5:
\longquote{%
It feels like quite a natural way you might approach this problem as a human. You're burrowing down into it more precisely and pushing your error around a little bit. I thought the transition of `we've moved the error from the log into the subtract [using log1p], now I know
how to deal with the error in a subtract as well' felt natural, ... since ... once we figure out it was going to be the subtract that was giving us trouble, then 
[we can use Herbie to rewrite successfully]. 
It gets there much faster, 
but it's cool that I also feel that 
I would have thought about going in a similar direction.}


In the survey, experts rated the interface's ability to generate ideas for
solving specific problems (item 4) with scores ranging from 5 to 7, with an
average of 5.8. The interface's effectiveness in evaluating the quality of ideas
quickly (item 5) was rated between 5 and 7, with an average of 6.4. These
relatively high ratings indicate that the experts found Odyssey helpful in
generating and evaluating ideas for improving floating-point expressions.

Users were able to use Odyssey to successfully generate 
  a variety of valid nontrivial new expressions for analysis, 
  both using an automated tool 
  (e.g. the way we expected users to solve Task 2) 
  and by themselves (P5 and P4). This was significantly different from our experience 
  in the earliest parts of the design process.
  The ability to send rewrites back to Herbie was a vital
  part of the solution generation process for the three experts who were able to complete Task 5.


\begin{table*}
  \centering
  {\renewcommand{\arraystretch}{1.25}
\begin{tabular}{||c|p{12cm}|c|c||}
    \hline
    \# & Survey Questions: & Results: & Average:\\
    \hline
    1 & "The workflow made sense to me and I was able to follow it." 
    & 5,  5,  6,  7,  7 & 6/7\\
    \hline
    2 & "This workflow matches my experience approaching real numerical analysis
    problems." & 4,  6,  6,  6,  6 & 5.6/7\\
    \hline
    3 & "The interface helped me identify or confirm specific problems with
    expressions." & 7,  7,  7,  7,  7 & 7/7\\
    \hline
    4 & "The interface allowed me to generate ideas for solving a specific problem." & 5,  5,  6,  6,  7 & 5.8/7\\
    \hline
    5 & "The interface let me evaluate the quality of ideas for rewritings quickly."
    & 5,  6,  7,  7,  7 & 6.4/7\\
    \hline
    6 & "It was easy to compare expressions in the interface." & 6,  6,  6,  7,  7 & 6.4/7\\
    \hline
    7 & "It was easy to mix together expressions from different sources in the interface." & 4,  5,  5,  6,  7 & 5.4/7\\
    \hline
    8 & "The interface let me focus on thinking about the problem at a high level." & 5,  6,  7,  7,  7 & 6.4/7 \\
    \hline
    9 & "I can think of ways to extend this workflow + interface to address numerical analysis problems that I have worked on." & 5,  6,  7,  7,  7 & 6.4/7 \\
    \hline
\end{tabular}}

  \caption {After completing the seven tasks, experts were asked to evaluate
 different aspects of the tool on a scale of 1 to 7. }
  \label{fig:survey-results}
\end{table*}

\subsubsection*{Third Stage: Tuning}

The third stage of our proposed workflow involves tuning expressions to further
optimize their accuracy and performance. To assess Odyssey's support for this
stage, we evaluated Task 7, as well as survey items 6 and 7, which inquired about
the interface's support for comparing and mixing different expressions.

Task 7 challenged experts to create a more accurate expression than
Herbie's best alternative for a given expression by combining different
solutions and fine-tuning the branch point. The task demonstrated that a human
can use Odyssey to outperform Herbie's internal heuristics when unique
requirements call for a tailored approach. 
After using the range zoom feature and noticing 
  Herbie's solution was still outperforming their solution 
  on a small region,
  P2 remarked, 
  \shortquote{So in this view, we can see that 
  we don't have quite the right number [for the branch point].} 
The expert then adjusted the branch point 
  based on the visual feedback.

In the survey, experts rated the interface's capacity to help them mix expressions from different sources (item 7) with
scores ranging from 4 to 7, with an average of 5.4. The interface's support for
comparing different expressions (item 6) was rated even more highly, at an
average of 6.4 (range from 6 to 7).

As we can see in the example above, 
  the especially high rating for comparison
  was likely a result of combining the ability 
  to plot the error for different expressions
  together with zooming to focus on getting feedback on specific regions.
A couple experts (P4, P5)
  mentioned wanting more support for combining expressions, 
  especially around conditional branches.
P4 explained that 
  an automated tool might be able to add guard conditions 
  where appropriate.

Finally, the experts appreciated the potential power 
  of mixing human and automated solutions, 
  with P3 commenting that 
  suggesting \texttt{log1p} and \texttt{hypot} to Herbie 
  felt similar to proof assistant tools where \shortquote{if you just add in an additional step on the way or an additional lemma... then it can actually nudge it over that threshold.}

In summary, the results from Task 7, along with the survey responses for items 6
and 7, provide evidence that Odyssey effectively supports tuning expressions for optimal accuracy and
performance. The interface enables users to mix expressions
and adjust coefficients while offering real-time feedback, streamlining the
tuning process and enhancing the overall quality of floating-point expressions.

\section{Discussion} \label{sec:discussion}
As the first expression rewriting workbench for the numerics community,
  Odyssey demonstrates how to build useful expert tools
  that enable users to more effectively search a design space. 
  Below, we discuss three insights that were key to Odyssey's design. 
  These insights serve as design principles that generalize to expert tools in other domains where users want to navigate a design space.

\paragraph{Expose heuristics, not states.}
  First, we found that exposing the internal exploration-focusing heuristics of the tool,
  rather than just the search states---%
  for Herbie, mainly the local error---helped users significantly,
  beyond its use in Herbie alone. 
By connecting this heuristic to other simpler metrics (like the input error plot), 
  users developed \textit{explanations} of the heuristic's value
  that helped them understand
  what was relevant about the search state---%
  for expression search,
  what subexpression was probably causing the error. 
By comparing the heuristic and their explanation across expressions, 
  users could check if the issue was solved, 
  even if the expression shape was too complicated for an automated tool to recognize.

  \paragraph{Give access to intermediate representations.}
  Second, we found that giving the user ownership over intermediate parts of the search
  made the tool much more useful. Doing so even allowed us to catch a bug in the underlying tool.
  A widely held belief among the HCI community is that higher levels of abstraction are more desirable for end-users. 
  Therefore, in an automated expert tool, it can seem natural to hide the middle of a search from the user to keep them working at a high level.
However, in our study, we found that users wanted to be able to see and control the search process. 
  Experts were particularly eager to introduce their own ideas and test assumptions.
In Odyssey, without building any separate tooling 
  except for a table that tracks candidates
  and synchronizes with visualizations of existing automated analyses, 
  users are able to explore a much broader space of possible solutions 
  in a way that was not possible with the original tool, 
  simply by letting a human manage search candidates.

\paragraph{Test expert workflows with relative novices}
Finally, we were able to identify the appropriate level of abstraction in Odyssey because of our own iterative design process that involved novices and experts. 
  Involving novices sensitized us to the foundational cognitive burdens experts had developed workarounds for. 
We realized that if our tool could not help a novice at least understand basic issues,
  it was likely too opaque for experts to use productively.
The local error plot, a key feature we would not have included without involving novices, ended up being the most praised feature by experts.


\paragraph{Applications to other domains with user-driven design space search}
While this paper focuses on floating-point analysis, 
the above key insights and findings suggest generalizable principles for user-driven design space search. 
The tool wrapped by Odyssey, Herbie, works in a way that should be 
familiar to anyone who has worked with a design space exploration tool
or classical AI search:
  it identifies a troublesome part of an expression, 
  applies algebraic rewrites or approximations to that part of the expression to obtain new expressions, 
  tests those expressions to see if they are worth exploring, 
  and finally merges the best options.

The shape of this process 
  matches the workflow we describe for an analyst 
  identifying and solving problems with an expression step-by-step 
  while tracking possible rewriting directions.
This search shape is used in tools across many domains,
  including in automated theorem provers, carpentry compilers,
  machining systems, and ASIC design space exploration tools. 
  Yet, expert tools in these domains do not apply the above three principles. 
  As a result, the tools remain difficult to use and error-prone. 
  We hypothesize that applying the principles will improve expert tools in other domains where users search a design space.

\subsection{Limitations and Future Work} 

A major limitation of our design process was the tight design loop 
  we had to maintain during development. 
While this was necessary to ensure
  we were building a system that would be useful to users, this meant
  we had to compromise on the polish of some features and 
  altogether avoid others which would take too long to implement or require
  disturbing many parts of the interface.
With more time, we plan to further improve the interface's layout and
  provide more structured expression editing support.



Of course, the main future work we have planned is to extend Odyssey to
  incorporate more analyses and sources of rewritings, including ideas like
  operation cost analyses and hardware-specific rewrites that were mentioned by
  the experts in our study.
Tools like PRECiSA~\cite{precisa} that already have an HTML-based analysis interface may be a good starting point 
  for testing these integrations. 

Floating-point experts were very appreciative of our work, and saw a variety of
  ways it could be extended to further support their particular areas of
  expertise.
These included ideas like adding support for multi-precision rewritings,
  incorporating operation cost analyses from Herbie and other tools, adding ways
  of helping human users simultaneously optimize at least 3 variables, and
  increasing support for splitting expressions into subregions and subexpressions
  based on domain-specific heuristics.


Odyssey also has clear potential application in floating-point education.
Several of our tasks asked users to explain to the interviewer potential
  problems with an expression using the interface, and both the experts and the
  novices in our formative study were able to point out areas of high error,
  select points, and zoom in to get a better look at problem regions to support
  diagnostic claims.
Odyssey has the potential to thrive in a classroom setting; it could be used by
  an instructor to show off how expression rewriting makes expressions more
  accurate or by students to explore and diagnose error sources an expression and
  try fixing them.
We plan to try applying Odyssey in an undergraduate class covering
  floating-point representations soon.

We are also excited by the explanatory potential offered by the incorporation
  of large language models (LLMs) like GPT.
We have found that available language models can, in fact, offer rewritings and
  generate plausible explanations for users, but they are prone to
  “hallucinating” and incorporating nonsensical logic, so their output must be
  validated before it is used.
With access to Odyssey's calculation and validation tools, 
  an LLM might be able to avoid these issues.

Finally, a major possible extension was brought up independently by two
  different participants, who commented that they would be very interested in
  plugging in additional visualizations showing actual output effects of errors
  for each expression.
For example, one participant has worked with expressions representing ellipses,
  and wanted to see how different kinds of error could lead to distortion of the
  ellipses.
Allowing for additional visualizations would be a major possible improvement,
  since it will help users understand whether the error they see on the error
  plot matters when code is compiled and run in practice.
If (as with ellipses) the output space can be mapped back to specific input
  values, combining output visualization with the error graph heatmap will let
  experts relate points with noticeable error in the actual output to the
  particular mathematical operation causing that error.

Overall, we are excited to see what floating-point experts and novices
  end up doing with Odyssey 
  and look forward to improving our support 
  for their work in the future.

\begin{acks}
We thank the many friends and members of the floating-point community who helped us work through the design of Odyssey, and Jon Froehlich and Joshua Horowitz 
for reading early drafts of this paper and offering feedback.
We also thank
our anonymous shepherds and reviewers for guidance
and valuable suggestions while preparing the final version of this
paper.
This work was supported by NSF award 901386 and the NSF Graduate Research Fellowship Program (GRFP).
This material is based upon work supported by the U.S. Department of Energy, Office of Science, Office of Advanced Scientific Computing Research, ComPort: Rigorous Testing Methods to Safeguard Software Porting, under Award Number DE-SC0022081. This work was also supported by the Applications Driving Architectures (ADA)
Research Center, a JUMP Center co-sponsored by SRC and DARPA.
\end{acks}

\balance{}

\bibliographystyle{ACM-Reference-Format}
\bibliography{references,odyssey}

\appendix







\section{Supplemental Material}

\subsection{Expert Study}

Below, we describe the procedure for our expert study of Odyssey.

\subsubsection{Introduction and Background (8 min)}
Our process began with an introduction and background session. 
This phase involved introductions 
  followed by a set of background questions 
  aimed at understanding the participant's experience 
  and usage habits around numerical analysis tools. 
We asked about the participant's years of experience, 
  when they last analysed 
  the error of a floating-point expression, 
  their typical workflow 
  for analysing high floating-point error expressions, 
  and their familiarity with the Herbie tool. 
If the participant was not a user of Herbie, 
  we sought to understand their reasons for not using it and 
  asked if there were ways they imagined Herbie
  fitting into their workflow.

\subsubsection{Tutorial (12 min)}
Following the introductory phase, 
  we gave the participant access to the Odyssey interface
  via Zoom and conducted a twelve-minute tutorial
  to familiarize them 
  with the Herbie interface. 
The tutorial demonstrated several features 
  using the expression $\sqrt{x + 1} - \sqrt{x}$ 
  for positive $x$. 
The features covered included 
  the specification of the expression being rewritten and ranges over which it must be accurate, 
  reading the error plots, local error identification, 
  selecting expressions from the rewriting table,
  expression editing, opening a new expression in a different tab, 
  and resampling on a different range. 
Throughout this tutorial, 
  we encouraged participants 
  to think out loud and provide feedback, 
  emphasizing our interest 
  in continuous interface improvement.

\subsubsection{Tasks (30-55 min)}
The next phase of our process was a task-oriented session 
  whose length depended on participant skill and availability. 
The tasks were designed to exercise different 
  parts of the interface and to reveal insights 
  about the participants' understanding 
  and ability to apply Odyssey for expression rewriting.
The tasks covered 
  identifying sources of error 
  in specific mathematical expressions, 
  using the Odyssey system 
  to find and recommend improvements, 
  evaluating the effectiveness of proposed solutions, 
  and identifying problems in automated solutions. 
For each task, 
  specific goals were set ahead of time in terms of interface usage 
  and problem-solving approach so we could decide 
  if Odyssey was able to meet the participant's need 
  and whether their usage represented a novel approach.
Here is the full list of tasks and usage goals:
\begin{enumerate}
    \item \textbf{Identify relevant sources of error in the Rust $asinh$ implementation.}
        \begin{itemize}
        \item The participant should be able to determine the cause of the error by clicking on two different points to see at least two local error graphs.
        \end{itemize}
        
    \item \textbf{Use Odyssey to find and recommend $hypot$.}
        \begin{itemize}
        \item The participant should submit $sqrt(x * x + 1)$ in a new tab, ask Herbie for rewritings, and obtain $hypot(1, x)$ or another solution.
        \end{itemize}
        
    \item \textbf{Determine whether the solution to task 2 is good enough.}
        \begin{itemize}
        \item The participant should be able to integrate the result from task 2 into the original expression and refer to the error plot to justify their answer.
        \end{itemize}
        
    \item \textbf{Identify problems with branches (regimes) in automated solutions.}
        \begin{itemize}
        \item The participant should be able to highlight areas of concern by clicking on points around 1, where higher error is shown.
        \end{itemize}
        
    \item \textbf{Use Odyssey to find and recommend a way of solving small $x$ with $log1p$.}
        \begin{itemize}
        \item The participant should be able to find a good solution for the entire range of positive $x$ values that doesn’t include branches using Herbie's suggestions.
        \end{itemize}
        
    \item \textbf{Determine trust in the expression.}
        \begin{itemize}
        \item The participant should be able to verify the expression's equivalence to the original by checking the expression derivation.
        \end{itemize}
        
    \item \textbf{Use Odyssey to create a branched solution.}
        \begin{itemize}
        \item The participant should be able to create a branched expression that outperforms Herbie’s solution.
        \end{itemize}
    \end{enumerate}

\subsubsection{Survey and discussion (10-15 min)}
In the final phase, 
  we conducted a Google Forms survey that lasted between 10 to 15 minutes.
  The survey questions and results can be found in Table \ref{fig:survey-results}.

\end{document}